\documentclass[doublecol]{epl2} 

\newcommand{\beq}{\begin{equation}}
\newcommand{\eeq}{\end{equation}}
\newcommand{\beqa}{\begin{eqnarray}}
\newcommand{\eeqa}{\end{eqnarray}}
\newcommand{\kvec}{{\bf k}}
\newcommand{\G}{{\cal G}}
\newcommand{\qvec}{{\bf q}}

\newcommand{\rvec}{{\bf r}}
\newcommand{\Avec}{{\bf A}}

\newcommand{\svec}{{\bf s}}

\title{Evidences for pairing of nearly-free quasiparticles from paraconductivity in layered superconducting cuprates}
\shorttitle{Evidences for pairing of nearly-free quasiparticles ...} 

\author{S. Caprara\inst{1} \and M. Grilli\inst{1} \and B. Leridon\inst{2} \and J. Vanhacken\inst{3}}
\shortauthor{S. Caprara \etal}

\institute{                    
  \inst{1} INFM-CNR, Unit\`a di 
Roma 1 and SMC Center, and Dipartimento di Fisica, Universit\`a di Roma 
``Sapienza'', piazzale Aldo Moro 5, I-00185 Roma, Italy \\
  \inst{2} Laboratoire de Physique Quantique - ESPCI/UPR5-CNRS, 10, Rue Vauquelin - 
75005 Paris - France \\
\inst{3}INPAC, Institute for Nanoscale Physics and Chemistry, Katholieke 
Universiteit Leuven, Celestijnenlaan 200 D, B-3001 Heverlee, Belgium
}

\abstract{
We revisit the Aslamazov-Larkin theory of paraconductivity in two dimensions,
to distinguish its universal features from the specific features of nearly-free
paired fermions. We show that both the numerical prefactor and the temperature 
dependence of the experimental paraconductivity in underdoped 
La$_{2-x}$Sr$_x$CuO$_4$ are only compatible with pairing of nearly-free 
fermionic quasiparticles. This conclusion is strengthened by the analysis of
paraconductivity data in the presence of a finite magnetic field, from which 
we extract a rather low value of the critical field $H_{c2}(T=0)$.
}

\begin{document}

\maketitle

\section{Introduction}
Layered superconducting (SC) cuprates are characterized by a pseudogap state in 
the underdoped region of the phase diagram, below a temperature $T^*$ which at 
low doping is much larger then the SC critical temperature $T_c$ and merges 
with it near optimal doping, where $T_c$ is maximum. A possible explanation 
relies on the formation of incoherent SC Cooper pairs below $T^*$, the modulus 
$|\Delta|$ of the SC order parameter acting as the pseudogap detected by 
various thermodynamical and transport measurements. Superconductivity is 
prevented by fluctuations of the phase of the order parameter, and develops 
only below $T_c$, where phase coherence is eventually established and the 
preformed pairs condense. The observation of a sizeable Nernst effect 
\cite{ong} and a strong diamagnetic response \cite{luli,wang} above $T_c$ have 
been interpreted in this sense 
\footnote{ This interpretation was recently questioned, and an alternative 
explanation in terms of Gaussian SC fluctuations was proposed, see, e.g., L. 
Cabo, {\it et al.}, Phys. Rev. Lett. {\bf 98}, 119701 (2007); N. P. Ong, 
{\it et al.}, {\it ibid.}, 119702 (2007). }. If this were the case, however, 
the most anisotropic cuprates [e.g., ${\rm Bi_2Sr_2CaCu_2O_{8+\delta}}$ 
(BSCCO)] should display an exponential temperature dependence in the 
enhancement of conductivity due to SC fluctuations at temperatures $T>T_c$ [the so-called 
{\it paraconductivity}], associated with vortical fluctuations, typical 
of a Kosterlitz-Thouless transition in two dimensions (2D) 
\cite{halperin}. Instead, it is well documented 
 that paraconductivity in all the families of cuprates is fully accounted for by the 
standard Aslamazov-Larkin (AL) theory \cite{AL,varlamov} based on Gaussian SC 
fluctuations, with the real and imaginary part of the SC order parameter $\Delta$ 
fluctuating around zero. While YBa$_2$Cu$_3$O$_{7-x}$ is less anisotropic and displays
the AL behavior characteristic of three-dimensional systems \cite{leridonPRL1}, all other compounds,
which have a more anisotropic structure, display the standard AL behavior for
two-dimensional systems (see, e.g., Refs. \cite{CGLL,brigitte} and references therein).
In particular recent experiments in underdoped 
La$_{2-x}$Sr$_x$CuO$_4$ (LSCO) recovered the normal state under strong magnetic 
field, thereby allowing for an unambiguous determination of paraconductivity 
\cite{brigitte},  
{\it leaving no room for a contribution of vortical phase fluctuations} over 
the broad temperature range relevant for the pseudogap. This result challenges 
the phase-fluctuation scenario raising the following issue: How stringent is 
the above conclusion based on the AL expression for paraconductivity? Within a general 
phenomenological Ginzburg-Landau (GL)  approach, we show that the AL functional form 
in 2D [$\propto (T-T_c)^{-1}$] is fairly general because ultimately stems from 
two general principles, namely gauge-invariance and the hydrodynamic form of 
the pair collective modes. On the contrary, the numerical prefactor is 
specific of the fermionic state and therefore provides valuable information on 
the microscopic state of the system. Specifically, we show here that the 
precise AL value of the paraconductivity coefficient stems from the assumption of fermions 
with very narrow spectral weight (i.e.,  nearly free fermionic quasiparticles).
This result is one of the two central points of this Letter and, together with experiments 
of Ref. \cite{brigitte}, which dictate the specific value and temperature 
dependence of this factor, clearly indicates that in underdoped LSCO 
fluctuations not only are Gaussian, but also arise from pairing of apparently 
{\it weakly-coupled} fermionic quasiparticles.

To challenge this quite surprising result, we present here new paraconductivity data in weak 
magnetic fields. We find that paraconductivity is still fully compatible with weakly paired 
quasiparticles and we also introduce the new concept of ``hidden'' 
critical field at zero temperature, $H_{c2}^G(0)$,
related to the Gaussian fluctuations only. Its value is remarkably lower than the 
one usually reported in the literature, strengthening our conclusion that paraconductivity 
is related to superconductivity due to  weakly-coupled quasiparticles.
This is the second remarkable point of this work.
These evidences of weakly-coupled quasiparticles, are surprising because their 
presence could hardly be guessed from the quite anomalous form of the normal 
state resistivity and is at odds with the broad spectral lines usually 
observed in photoemmission experiments in cuprates \cite{damascelli}. 
Our aim is not to solve this apparent contradiction, but rather to draw 
attention to this feature. To extract all information from the data, we 
preliminarly revisit the theoretical derivation of the Gaussian theory, 
putting precise bounds to the meaning and generality of the 2D AL expression.

\section{Gauge invariant hydrodynamic description of paraconductivity}
A superconductor can be described within a generic model of fermions coupled 
by a $\lambda$-wave pairing interaction (most frequently $s$- or $d$-wave have 
been considered for singlet superconductors). As customary, by integrating out 
the fermions one derives an effective action for the pair field 
$\Delta(\rvec,\tau)$ (here $\rvec$ is the coordinate vector and $\tau $ is the 
imaginary time within the finite-temperature formulation). The quadratic 
(Gaussian) part of the resulting action is
\beq
{\cal S}_G= \int_0^\beta d\tau \int d^D{\qvec}\, \Delta^*(\qvec,\tau) 
\left[a+ C\qvec^2 +\gamma\partial_\tau\right] \Delta(\qvec,\tau),
\label{effaction}
\eeq
where $D$ is the space dimensionality, $\Delta(\qvec,\tau)$ is the Fourier 
transform of $\Delta(\rvec,\tau)$ with respect to $\rvec$, and $\qvec$ is the 
corresponding wavevector. Whereas the explicit expressions of the coefficients 
$a$, $C$, and $\gamma$ depend on the details of the microscopic model, e.g., 
the pairing symmetry and the fermionic density of states (DOS), Eq. 
(\ref{effaction}) holds generically whenever a hydrodynamic description for 
the pair field is adequate, and is indeed phenomenologically adopted in the 
time-dependent Ginzburg-Landau approach \cite{varlamov}. In this Letter we 
consider Gaussian fluctuations above $T_c$, and keep only the action 
(\ref{effaction}), discarding higher-order terms. In the GL approach, one may 
conventionally take $\gamma=\gamma_{GL}=1$, which amounts to rescale $\Delta$ 
so that its equation of motion is the Schr\" odinger equation. Thus, in the 
Gaussian approximation, physical quantities only depend on two parameters, the 
{\em mass} $a_{GL}\equiv a/\gamma$ and the {\em stiffness} 
$C_{GL}\equiv C/\gamma$.

The pair field, with a charge $2e$, is coupled to a spatially uniform 
electromagnetic field $\Avec(\tau)$ taking $\qvec\to\qvec -2e \Avec(\tau)$, 
as dictated by gauge invariance. The AL contribution to the current-current 
response, and hence to paraconductivity, is associated with the current density 
$4eC\qvec\,\Delta^*(\qvec,\tau)\Delta(\qvec,\tau)$, and the prefactor in the 
current vertex can be identified with the stiffness. Under the assumption of a 
{\it gauge-invariant hydrodynamic description} for the SC pair fluctuations, 
the above arguments hold {\it irrespective of the Fermi-liquid or 
non-Fermi-liquid character of the normal state}. Of course, any microscopic 
derivation of Eq. (\ref{effaction}) must obey gauge invariance. In the case of 
strongly interacting fermions such a derivation is overwhelmingly difficult 
and beyond the scope of this Letter. On the other hand in the following 
section, we provide an example of the current-stiffness relation in the case
of weakly-coupled fermions. 

\section{Weak-coupling microscopic derivation}
Our treatment closely follows the gauge-invariant approach of Ref. 
\cite{ambegaokar}. We start from a Baym-Kadanoff functional (i.e., the 
microscopic equivalent of the GL functional) and obtain the paraconductivity by insertion of 
current vertices. For weakly-coupled fermions one can adopt the Baym-Kadanoff functional 
shown in Fig. 1($a$). 

\begin{figure}
\onefigure[scale=0.3]{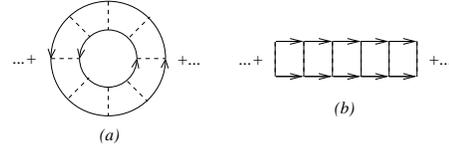}
\caption{Typical diagram for the Baym-Kadanoff functional ($a$) and $\mathcal T$-matrix
propagator of Gaussian fluctuations ($b$) adopted in this Letter. Dashed and 
solid lines represent, respectively, the pairing interaction and the fermion 
propagator (see text).}
\label{fig1}
%
\end{figure}

For definiteness we assume a separable potential
$V(\kvec,\kvec')=V w_\lambda(\kvec)w_\lambda(\kvec')$ of strength $V$,
promoting $\lambda$-wave pairing [in cuprates, e.g., $d$-wave, with 
$w_d=\cos(k_x)-\cos(k_y)$]. A weak-coupling $\mathcal T$-matrix approximation 
yields the pair propagator of Fig. 1($b$), i.e., the inverse of the 
coefficient of the action (\ref{effaction}), 
\beq
{\cal K}_\lambda(\qvec,\omega_\ell)=\frac{1}{V^{-1}-\Pi_\lambda(\qvec,\omega_\ell)}
\approx\frac{1}{a_\lambda+C_\lambda \qvec^2+\gamma_\lambda|\omega_\ell|},
\eeq
with the 
$\tau$ variable Fourier-transformed into the Matsubara frequency $\omega_\ell$. 
The $\lambda$-wave particle-particle bubble is
\beq
\Pi_\lambda(\qvec,\omega_\ell)\equiv T\sum_{\kvec,\varepsilon_n} 
w_\lambda^2(\kvec)\,\G(\kvec +\qvec,\varepsilon_n+\omega_\ell)
\G(-\kvec,-\varepsilon_n),
\eeq 
$\G(\kvec,\varepsilon_n)\equiv(i\varepsilon_n-\xi_\kvec)^{-1}$ is the fermion 
propagator, and $\xi_\kvec$ is the fermion dispersion. An expansion of 
$\Pi_\lambda(\qvec,\omega_\ell)$ at small $\qvec$ and $\omega_\ell$ yields, 
respectively, $C_\lambda$ and $\gamma_\lambda$. The mass
$a_\lambda\equiv V^{-1}-\Pi_\lambda(0,0)$ linearly vanishes at $T=T_c$.

\begin{figure}
\onefigure[scale=0.3]{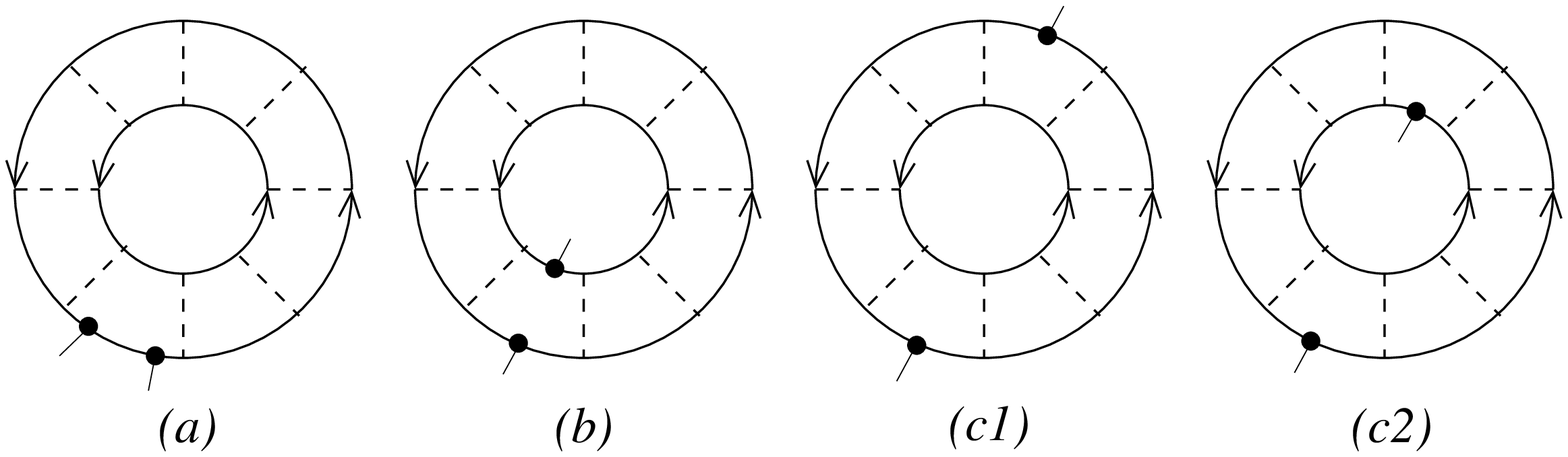}
\caption{Diagrams of the current-current response functions generated from 
the Baym-Kadanoff functional of Fig. 1($a$): DOS correction ($a$), Maki-Thompson vertex 
correction ($b$), and AL contributions ($c1$, $c2$). The full circle with a 
thin line represents a current-vertex insertion (see text).}
\label{fig2}
%
\end{figure}

The insertion of two current vertices in the diagrams of Fig. 1($a$) yields the 
current-current correlation functions \cite{ambegaokar} shown in Fig. 2. The 
diagrams of Figs. 2($c1$) and 2($c2$) give the AL contributions, once the 
ladder resummation of Fig. 1($b$) is adopted for the pair propagator. These 
contributions are different from the others, as they vanish if the fermionic 
loops with one current-vertex are evaluated for zero frequency and momentum of 
the pair propagators, due to the vector character of the current vertex. 
The first non-zero contribution to each loop is $\tilde C\qvec$ \cite{AL},
where $\tilde C$ is a constant prefactor. Gauge invariance imposes a definite 
relation between $\tilde C$ and the stiffness $C$. This relation is enforced by 
a Ward identity, which can be derived from the Baym-Kadanoff functional, and
to first order in the momentum difference $\svec$ reads
\beqa
&& {\cal K}_\lambda^{-1}(\qvec+\svec,\omega_\ell)-
{\cal K}_\lambda^{-1}(\qvec,\omega_\ell)=\nonumber \\
&&T \sum_{\kvec,\varepsilon_n}
w_\lambda^2(\kvec)\G(\kvec,\varepsilon_n)\G(\kvec,\varepsilon_n)
\G(-\kvec+\qvec,-\varepsilon_n)\,{\mathbf v}_\kvec\cdot\svec, ~~~~
\eeqa
where ${\mathbf v}_\kvec\equiv \partial_\kvec\xi_\kvec$ is the fermion velocity,
acting as a current vertex in the fermion loops. The direct calculation in the weak-coupling limit 
yields indeed $2C=\tilde C$, where the factor of 2 stems from the $2e$ charge 
of the pair field. 

\section{What can be inferred from observation of AL paraconductivity}
The identification of the coefficient $\tilde C$ of the AL current vertex with 
the stiffness $C$ is the reason why the AL paraconductivity in 2D assumes an expression 
which is independent of $C$. Indeed, the AL current-current response reads 
\cite{AL,varlamov}
\beqa
&&\delta \chi_{AL}(\Omega_n)= 4e^2T\sum_{\omega_\ell}\int d^D\qvec\,
\frac{1}{a_{GL,\lambda} +C_{GL,\lambda}\qvec^2 +|\omega_\ell|}\nonumber\\
&& \times\frac{1}{a_{GL,\lambda}+C_{GL,\lambda}\qvec^2+|\omega_\ell+\Omega_n|}
\,C_{GL,\lambda}^2 \qvec^2,\label{chiJJ}
\eeqa
where the dependence on $\gamma_\lambda$ was eliminated in the GL spirit, 
introducing the two independent parameters 
$a_{GL,\lambda}\equiv a_\lambda/\gamma_\lambda$, 
$C_{GL,\lambda}\equiv C_\lambda/\gamma_\lambda$, as discussed above. In the 
classical limit the sum over $\omega_\ell$ is dominated by the term 
$\omega_\ell=0$. After the analytic continuation $i\Omega_n\to\omega+i0^+$, the 
AL paraconductivity is found as $[\mathrm{Im}\delta\chi_{AL}(\omega)/\omega]_{\omega\to 0}$. In 
2D the change of variables $C_{GL,\lambda}\qvec^2\to x$ makes $C_{GL,\lambda}$ 
disappear, yielding the well-known result \cite{AL}
\beq
\delta \sigma_{AL}(\varepsilon)=
\frac{e^2 }{2\pi \hbar d }\frac{T_c}{a_{GL,\lambda}}\equiv
\frac{e^2}{16\hbar d \varepsilon},\label{ALgeneric}
\eeq
where $d$ is the interlayer distance, translating the 2D result into the paraconductivity of 
a layered system, and $\varepsilon\equiv\pi a_{GL,\lambda}/(8 T_c)$ is the 
dimensionless mass. Eq. (\ref{ALgeneric}) stems from the assumption of a 
gauge-invariant hydrodynamical description for the Gaussian pair fluctuations, 
which in 2D imposes the independence from $C_{GL,\lambda}$, and is thus generic 
for 2D Gaussian fluctuations 
\footnote{Actually the functional form of  Eq. (\ref{ALgeneric}) is even more general
since it also holds for 
Kosterlitz-Thouless phase fluctuations with $\varepsilon$ exponentially vanishing at $T_c$ 
\cite{halperin,demlersachdev}.}.

Since we aim to extract as much physical content as possible from the fitting
of experimental data with Eq. (\ref{ALgeneric}), we now detail the specific 
value of the coefficients in the various physical situations. All information 
on the microscopic physical properties is contained in $\varepsilon$. As soon 
as the fermion DOS changes with temperature (e.g., with the opening of a 
pseudogap) one may wonder how this is reflected in the temperature dependence 
of $a_{GL}$ for the various pairing regimes. In a BCS model of weakly-coupled 
fermions, the explicit calculation of the particle-particle bubble 
$\Pi_\lambda$ can be carried out, yielding
\begin{eqnarray*}
\gamma_\lambda &=&-\sum_{\kvec} w_\lambda^2(\kvec)\int dz\, A(\kvec,z)
A(\kvec,-z)\,\partial_z f(z) \nonumber\\
a_\lambda &=&V^{-1} -\sum_\kvec w_\lambda^2(\kvec) \int dy\,dz\, 
A(\kvec,y) A(\kvec,z)\mathcal R(y,z)\nonumber  
\end{eqnarray*}
where $f(z)$ is the Fermi function, $\mathcal R(y,z)=[1-f(y)-f(z)]/(y+z)$,
and $A(\kvec,z)$ is the fermion spectral function. If this latter is narrower 
than $\partial_z f(z)$, it can be replaced by $\delta(\xi_\kvec-z)$. In this 
case, a symmetry-dependent weighted DOS 
${\mathcal N}_\lambda\equiv\sum_{\kvec}w_\lambda^2(\kvec)\delta(\xi_\kvec)$
appears, generalizing the standard $s$-wave expressions of the
$\gamma$ and $a$ coefficients \cite{varlamov}. This factor enters both in 
$\gamma_\lambda$ and $a_\lambda$, and disappears in 
$a_{GL}\propto a_\lambda/\gamma_\lambda$ leaving the paraconductivity unaffected by the $T$ 
dependence of the DOS. It is important to recognize that this result follows
from the narrow spectral density of the fermions entering the Cooper channel,
and is no longer valid if the spectral density is broad. This suggests that 
the absence of any additional temperature dependence is the specific signature 
of paraconductivity from weakly-paired nearly-free quasiparticles. In any case, the  
numerical prefactor relating  $a_{GL}$ to $(T-T_c)$ is model dependent and the 
standard result $\varepsilon=\log(T/T_c)$ is a specific signature of the BCS 
weak-coupling limit. Therefore, a {\it pure} AL contribution [with the 
specific $e^2/(16\hbar d)$ prefactor and $\varepsilon =\log(T/T_c)$] is hardly 
mistaken and is a clear indication of nearly-free fermions being 
{\it the only} carriers responsible for paraconductivity via the formation of fluctuating 
weakly-coupled Cooper pairs, independently of their DOS and its possible 
temperature dependence. This observation is the crucial point of our 
theoretical analysis: from the data shown below it will allow us to infer that 
electrons giving rise to paraconductivity in LSCO behave {\it as if} they were forming 
Cooper pairs of weakly coupled nearly-free quasiparticles. We now apply these 
theoretical conclusions to the data obtained in underdoped LSCO.

\section{Evidence of nearly-free quasiparticle pairing} 
The resistance of 
several LSCO samples at different dopings has been recently measured as a 
function of $T$ with and without strong magnetic fields $H$ \cite{brigitte}. 
The complete destruction of the SC state at $H=47$ T uncovers a highly unusual 
normal state with a resistivity well reproduced, over an extended temperature 
range below 200 K, by the superposition of a linear and a logarithmic term 
$\rho_N(T)\equiv\rho(T,H=47\,{\rm{T}})=AT-B\ln(T/T_0)$, which naturally 
introduces a temperature scale at which a minimum in the resistivity occurs in 
underdoped cuprates under strong magnetic fields \cite{brigitte,boebi}. For a 
sample with $x=0.09$ and $T_c=19.0$ K our fit gives $A=7.54$ $\mu\Omega$cm/K, 
$B=490$ $\mu\Omega$cm, and $T_0=80.3$ K. We propose no explanation or 
hypothesis for this unusual normal state and rather focus on the SC state 
appearing when $H$ is reduced. Following Ref. \cite{brigitte}, we define the paraconductivity 
as $\delta\sigma(T)\equiv\rho^{-1}(T,H=0)-\rho_N^{-1}(T),\label{deltasigma}$
and report the results  in Fig. 3 (black dots) as a function of 
$\varepsilon\equiv\ln(T/T_c)$, in comparison with the 2D AL result in the BCS 
limit (solid line). Despite the unusual $\rho_N$, $\delta\sigma(T)$ is very 
well described by the standard AL expression {\it with the pure BCS 
coefficients, without fitting parameters}. Most importantly, we find that not 
only the temperature dependence is clearly linear in $\varepsilon^{-1}$, but 
even the numerical prefactor is that of the weak-coupling theory for 
nearly-free fermions, within error bars of less than 5\%.
Since the paraconductivity diverges at $T_c$, uncertainties in the determination of 
$\rho_N$ are rather immaterial for $T\approx T_c$ and our finding is quite 
robust. The contribution of Gaussian fluctuations to paraconductivity spreads over a broad 
temperature range, $T-T_c \sim T_c$, similarly to what found in underdoped 
BSCCO \cite{CGLL}, where however the need to guess the reference normal state 
made the analysis much less stringent.

Rewriting $\varepsilon=(\xi_0/\xi)^2$, and assuming $\xi_0\sim 20$ \AA, we can 
estimate the coherence length $\xi$ of the Gaussian fluctuations. Even for 
$\varepsilon\approx 0.01$, i.e., $T\approx 1.01T_c$, we find 
$\xi\sim 10\xi_0\sim 200$ \AA, which is much smaller than the value estimated 
for Kosterlitz-Thouless vortical phase fluctuations in magnetometry experiments in BSCCO 
\cite{luli}. This discrepancy can hardly be due to the different materials, 
because paraconductivity experiments in BSCCO \cite{CGLL} give values of $\xi$ consistent 
with those obtained here for LSCO. 

\begin{figure}
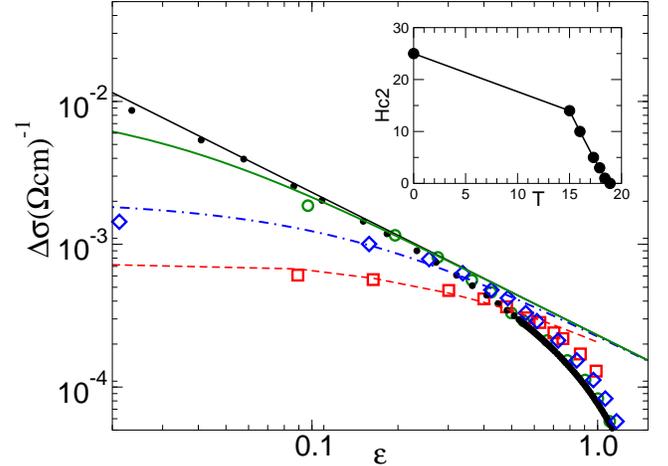

\onefigure[scale=0.32]{BrigitteFig3-4-c.eps}
\caption{(Color online) Comparison between the theoretical Gaussian paraconductivity
$\delta\sigma (T,H)$ \cite{varlamov} (lines) and the experimental data 
(symbols) taking an interlayer distance $d=6.6$ \AA\ [data at $H=0$ T 
(black dots), should be compared with theoretical result, Eq. 
(\ref{ALgeneric}), (black solid line)]. $H=1$ T (green solid line and 
circles), $5$ T (blue dot-dashed line and diamonds), and $14$ T (red dashed 
line and squares). Inset: Gaussian critical temperatures vs $H$ and estimated 
$H_{c2}(T=0)$ (see text).}
\label{fig34}
%
\end{figure}

We now focus on new data showing the gradual suppression of Gaussian 
fluctuations for small-to-moderate $H$. Since dissipating vortices, introduced 
by the magnetic field, largely contribute to the resistivity, the Gaussian paraconductivity 
is difficult to extract. Nevertheless we tested the 2D AL theory at finite $H$ 
using the expression reported in Ref. \cite{varlamov}. This attempt is 
obviously meaningful only if the critical temperature {\it in the absence of 
dissipating vortices}, $T_c^{G}(H)$, does not fall deeply into the 
vortex-dissipation regime. In Fig. 3 we report our results. The choice of 
$H_{c2}^G(T=0)$ and of $T_c^{G}(H)$ is made to optimize the agreement with the 
data. For $H=1,\, 5,\, 14$ T we find $T_c^{G}=18.4,\, 17.3,\, 15.0$ K, 
respectively (see the inset in Fig. 3), which are substantially larger than the 
experimental $T_c(H)$, determined by vortex dissipation. Therefore our analysis 
reliably indicates that 2D Gaussian fluctuations persist under substantial 
magnetic fields. We find $H_{c2}^G(T=0)=25$ T, which is much lower than the 
values at which superconductivity is actually destroyed and usually
reported for LSCO at $x=0.09$ \cite{hc2}. However, this value is 
estimated from the weak-coupling expression of Ref. \cite{varlamov}, 
and therefore should be 
interpreted as {\it the critical field of a system in which the physics of
vortices is absent and only Gaussian fluctuations play a role}. The success 
of our fitting procedure is based on the existance of a regime where paraconductivity
is due to Gaussian fluctuations and would completely fail if only preformed pairs
with vortical excitations were present.

\section{Conclusions}
In this work we started from the preliminary remark that 
AL paraconductivity is ubiquitously observed in cuprates.
This lead us to reexamine the theoretical grounds of AL theory
in order to fully ascertain the physical implication of this
phenomenological remark. We showed that under general conditions  (i.e., 
gauge invariance and hydrodynamics) 2D paraconductivity is independent of the fluctuation 
stiffness, and depends on a single parameter, the dimensionless mass 
$\varepsilon$, which contains all informations on the specific character of 
the paired fermions. Therefore the robustness of the AL functional form in 2D stems from general
physical principles, but the specific numerical prefators may shed light on the
nature of the paired fermions. In particular 
we showed that paraconductivity of the AL functional form with the
precise and specific AL prefactors can only be due to weakly-bound nearly-free fermions.

As far as the experimental part of our work is concerned, we concentrated 
on LSCO only because the new data in strong magnetic field 
allowed for the unambiguous determination of the reference normal state, 
but our analysis applies to all families of cuprates.
Thus we investigated the experimental paraconductivity in underdoped 
LSCO showing that it is fully accounted for by Gaussian fluctuations, both in 
the absence and in the presence of a magnetic field. The supporting 
theoretical analysis allows to conclude i) that within the experimental
errors, in paraconductivity there is no room for 
contributions due to vortical phase fluctuations, which seem instead 
to be present in other experimental  quantities \cite{ong,luli,wang}. Moreover
ii) the specific value of the numerical prefactor and the 
temperature dependence of the experimental dimensionless mass indicate that, 
despite the very anomalous normal state uncovered by the magnetic field, 
Gaussian fluctuations arise from the pairing of nearly-free fermionic 
quasiparticles. This would agree with the recent observation of a (small) 
Fermi surface of nearly-free electrons in underdoped YBa$_2$Cu$_3$O$_{6.5}$ 
\cite{proust}. This  indication of pairing of weakly coupled quasiparticles,
 whose presence can hardly be guessed from other physical properties of the cuprates,
is perhaps the most surprising and intriguing result of our analysis.

One might speculate that the weakly bound 
pairs probed by paraconductivity coexist with more tightly bound pairs related to the 
vortical phenomenology. This coexistance, already implicit in a previous 
analysis of a two-gap model \cite{twogap}, could also be consistent with recent 
observations of different gap scales \cite{letacon}.

In this work we also carry out, both experimentally and theoretically, a new analysis by 
investigating paraconductivity in magnetic field, introducing the new concept 
of "hidden" critical field related to weakly-bound 
pairs, which is usually masked by the vortex physics and which
rules the destruction of Gaussian fluctuations around $T_c$.


\acknowledgments
We are indebted with C. Castellani, C. Di Castro, J. Lesueur and A. Varlamov 
for interesting discussions. S.C. and M.G. acknowledge financial support from 
MIUR-PRIN 2005 - prot.\ 2005022492. S.C., M.G. and B.L. acknowledge support 
from CNRS PICS \#3368. B.L. also acknowledges the ESF for support through the 
THIOX short visit grant number 1081.

\end{document}